\documentclass[aps,prd,twocolumn,superscriptaddress,showpacs,balancelastpage]{revtex4}
\usepackage{graphicx,subfigure}

\input epsf

\def\aj{Astron. J. }

\font\FermiSmallfont=cmssq8 scaled 1200
\def\LANLppthead#1#2{
\null 
\begin{center}\vskip -1.0truein{\hbox to 7.5truein {
\hfill
\vbox to 1in {\vfill \FermiSmallfont
              \hbox{#1}
              \hbox{#2}
              \vfill}
}}\vskip-0.0truein\end{center}}
\def\be{\begin{equation}}
\def\ee{\end{equation}}
\def\bea{\begin{eqnarray}}
\def\eea{\end{eqnarray}}

\def\GeV{\,{\rm GeV}}

\def\fun#1#2{\lower3.6pt\vbox{\baselineskip0pt\lineskip.9pt
  \ialign{$\mathsurround=0pt#1\hfil##\hfil$\crcr#2\crcr\sim\crcr}}}

\def\fun#1#2{\lower3.6pt\vbox{\baselineskip0pt\lineskip.9pt
  \ialign{$\mathsurround=0pt#1\hfil##\hfil$\crcr#2\crcr\sim\crcr}}}

\def\Vmax{V_{\rm max} }
\def\rmax{r_{\rm max}}
\def\rhos{\rho_s} 
\def\rs{r_s} 
\def\Aeff{A_{\rm eff}}
\def\texp{t_{\rm exp}}
\def\BV{B_{\rm V}} 
\def\BG{B_{\rm G}}

\def\s{{\rm s}}
\def\cm{{\rm cm}}
 \def\Eth{E_{\rm th}}
 \def\Mchi{M_{\chi}}

\def\Vmax{V_{\rm max}} 
\def\rmax{r_{\rm max}}

\begin{document}

\LANLppthead{LA-UR 07-4969}{astro-ph/yymmnnn}

\title{The Most Dark Matter Dominated Galaxies: Predicted Gamma-ray Signals 
\\ from the Faintest Milky Way Dwarfs}

\author{Louis E. Strigari}
\email{lstrigar@uci.edu}
\affiliation{Center for Cosmology, Department of Physics and Astronomy, 
University of California, Irvine, CA 92697,
 USA}

\author{Savvas M. Koushiappas}
\affiliation{Theoretical Division,
  \& ISR Division, MS B227, Los Alamos National Laboratory, Los
  Alamos, NM 87545, USA}

\author{James S. Bullock}
\affiliation{Center for Cosmology, Department of Physics and Astronomy, 
University of California, Irvine, CA 92697,
 USA}

\author{Manoj~Kaplinghat} 
 \affiliation{Center for Cosmology, Department of Physics and Astronomy,  University of
California, Irvine, CA 92697, USA}

\author{Joshua D. Simon}
\affiliation{Department of Astronomy, California Institute of Technology, 
1200 E. California Blvd., MS105-24, Pasadena, CA 91125 USA}

\author{Marla Geha}
\affiliation{National Research Council of Canada, Herzberg Institute of Astrophysics,
5071 West Saanich Road, Victoria, BC V9E, 2E7, Canada}

\author{Beth Willman} 
\affiliation{Harvard-Smithsonian Center for Astrophysics, 60 Garden St. 
Cambridge, MA, 02138}

\date{\today}
\begin{abstract} 
We use kinematic data from three new, nearby, extremely low-luminosity 
Milky Way dwarf galaxies (Ursa Major II, Willman 1, and Coma
Berenices) to  constrain the properties of their dark matter halos,
and from these make predictions for the  $\gamma$-ray flux from
annihilation of dark matter particles in these halos. We show that
these $\sim 10^3$ L$_\odot$ dwarfs are the most dark matter dominated
galaxies in the Universe, with total masses within 100 pc in excess of
$10^6$ M$_\odot$. Coupled with their 
relative proximity, their large  masses imply that they should have
mean $\gamma$-ray fluxes comparable to or greater than any  other
known satellite galaxy of the Milky Way. Our results are robust to
both variations of  the inner  slope of the density profile and the
effect of tidal interactions. The fluxes could be boosted by up to two
orders of magnitude if we include the density enhancements caused by
surviving dark matter substructure. 

\end{abstract} 


\maketitle
\section{Introduction} 
The census of the Local Group has changed dramatically in the last few
years. Prior to the turn of the century, there were only eleven known
satellite galaxies of the Milky Way (MW), with a discovery rate of
roughly one new Local Group satellite per decade \cite{Mateo:1998wg}.   
However, the Sloan Digital Sky Survey (SDSS) has been able to uncover 
a population of extremely  low-luminosity  satellite galaxies, which has roughly 
doubled the number of known satellites \citep{discovery,Zucker:2006bf,Belokurov:2006ph}. 
Determining how these new satellites fit in a given model for dark matter and 
cosmology presents a very exciting theoretical challenge.

The Cold Dark Matter (CDM) model predicts the existence of hundreds of
MW satellites that are expected to host galaxies at the faint
end of the luminosity  function \cite{missingsatellites}. The ability
of gas to cool and form stars in these low mass dark matter halos 
depends on a number of complex physical processes, such as  supernova
feedback, the photoionizing background, as well as mass loss
due to tidal interactions \cite{dsp}. Despite the broad range of
observed luminosities, the dark matter masses for all of the pre-SDSS
satellites are  constrained to within relatively narrow range,
approximately $\sim [1-6] \times 10^7$ M$_\odot$ within their inner
600 pc  \cite{Walker:2007ju,Strigari:2007ma}.  
Understanding this strong luminosity bias at the low mass end is
crucial to deciphering the formation of these dwarf spheroidal (dSph) 
galaxies as well as to constraining the nature of dark matter.

In this paper we show that three new and nearby members of the Local Group
discovered by the SDSS (Willman 1, Coma Berenices, and Ursa Major II)
are likely to have masses comparable to their 
more luminous counterparts. Initial estimates have already shown that
these galaxies have mass-to-light  ratios similar to or larger than
the pre-SDSS dwarfs \cite{Martin:2007ic,Simon:2007dq}.  
With luminosities more than two orders of magnitude less than the
pre-SDSS dwarfs,  these new satellites are not only interesting in
the context of galaxy formation at the  lowest-mass scales, but also
for indirect dark matter detection. The new dwarfs are very faint, but 
they contain large amounts of dark matter and are located quite nearby, 
making them ideal sites to search for signals of dark matter annihilation.

Current and future observatories, including  
space-based experiments, such as GLAST \cite{GLAST}, as well as a suite of 
ground-based  Cerenkov detectors, such as STACEE \cite{Hanna:2002bf}, 
HESS \cite{Hofmann:2003kx}, MAGIC \cite{Cortina:2005pt}, 
VERITAS \cite{Weekes:2001pd}, CANGAROO \cite{Yoshikoshi:1999rg}, \& 
HAWK \cite{Sinnis:2005un}, will search for the signal 
of $\gamma$-rays from dark matter annihilations. 
The prospects for $\gamma$-ray detection from 
dark matter in well-known MW satellites with these observatories has been the
subject of many previous studies 
\cite{previousdwarfs,Bergstrom:2005qk,Strigari:2006rd}.
All of these systems are interesting targets not only because of their
large  mass-to-light ratios, but also because they are expected to
have very  low intrinsic  $\gamma$-ray emission. This is in contrast
to the situation at the Galactic center,   where astrophysical
backgrounds hinder the prospects of extracting the signal from  
dark matter annihilation \cite{galacticcenter}. Moreover, the known
location of MW  satellites makes a search of dark matter annihilation
well-defined, unlike the  search of completely dark substructure,
which would rely on  
serendipitous discovery \cite{microhalos,Koushiappas:2006qq,Diemand:2006ik}. 

From the mass modeling of the dark matter halos, we provide the 
first determination of the $\gamma$-ray
signal from dark matter from Ursa Major II,  Willman 1, and Coma
Berenices (Coma  hereafter). These galaxies provide promising targets for $\gamma$-ray
detection for three  reasons: 1) they are the among the closest dark matter dominated systems,  
2) they are expected to be  free from intrinsic $\gamma$-ray emission, 
and 3) present data on their stellar kinematics suggest that their
dark matter halos are as massive as the more well-known population of
MW satellites.  

This paper is organized as follows. In section~\ref{sec:theoreticalmodeling}, 
we review  the theoretical modeling of dwarf dark matter halos and the 
calculation of the $\gamma$-ray flux. In section~\ref{sec:likelihoodsection}, 
we present the likelihood function for determining the flux, and outline the 
theoretical priors in the modeling. In sections~\ref{sec:results}
and~\ref{sec:conclusions} we present the results and  discussions. 

\section{Theoretical Modeling} 
\label{sec:theoreticalmodeling}

When modeling the stellar distribution of a dwarf galaxy, it is important to 
determine the effect of external tidal forces on the dynamics of the system. 
For the MW satellites we study, we can obtain an estimate of 
the external tidal force by comparing its magnitude to the internal 
gravitational force. The internal gravitational force is $\sim \sigma^2/R$ and the 
external tidal force from the MW potential is $\sim (220 \, {\rm km}/{\rm s})^2R/D^2$, 
where $D$ is the distance to the dwarf from the center of the MW, and
$220 \, {\rm km}/{\rm s}$ is the MW rotation speed at the distance of
the dwarf. 
The MW satellites are characterized by scale radii of $R \sim
10-100$ pc and velocity dispersions $\sigma \sim 5-10$  km s$^{-1}$.
For a typical distance of $D \sim 40$ kpc, the internal gravitational forces are thus 
larger by $\sim 100$. 

An additional estimate of tidal effects can be obtained by comparing 
the internal crossing times of the stars in the galaxy to the orbital
time scale of the system in the external potential of the host. 
For a galaxy with scale radii and velocity dispersions given
above, 
an estimate of the crossing time is given by  $R/\sigma \sim 1-20$
Myr. Assuming a rotation speed of $\sim 220$ km s$^{-1}$ at the 
distance of these dSphs ($\sim 40$ kpc), their orbital
time scale  in the MW potential is $\sim$ Gyr. 

From the above estimates we conclude that {\em it is highly 
unlikely that these galaxies are presently undergoing significant tidal stripping.}
These galaxies may have been tidally stripped before (for example, if their orbit 
took them closer to the center of the MW), however the stellar core that
has survived is faithfully tracing the local potential \cite{tides,Penarrubia:2007zx}. 
Of the galaxies we consider, only Ursa Major II shows strong evidence of past 
tidal interaction, as it is located on the same great circle as the Orphan Stream 
discovered by SDSS~\cite{Zucker:2006bf,Belokurov:2006ph}.
Thus we can proceed ahead with confidence in modeling the surviving stellar cores
as systems in dynamical equilibrium.  

Line-of-sight velocities are widely used to determine the
properties of the  dark matter halos of dSphs 
\cite{Lokas:2004sw,Mashchenko:2005bj,Gilmore:2007fy} assuming
spherical symmetry. 
For a system in dynamical equilibrium, the spherically symmetric Jeans
equation gives the  stellar line-of-sight velocity dispersion at a
projected radius, $R$,  from the center of the galaxy,  
\be
\sigma_{t}^{2}(R) = \frac{2}{I(R)} \int_{R}^{\infty} 
\left ( 1 - \beta \frac{R^{2}}{r^2} \right )
\frac{\rho_{\star} \sigma_{r}^{2} r}{\sqrt{r^2-R^2}} dr, 
\label{eq:sigma}
\ee 
where the three-dimensional velocity dispersion, $\sigma_{r}$ is obtained 
from 
\be 
r \frac{d(\rho_{\star} \sigma_r^2)}{dr} =  - \rho_{\star}(r) V_c^2(r)
        - 2 \beta(r) \rho_{\star} \sigma_r^2.
\label{eq:jeans}         
\ee 
Here $\beta$ is the stellar velocity anisotropy, and $\rho_\star(r)$
is the density profile for the  stellar distribution, which is
obtained from the projected stellar distribution, $I(R)$. 
The stellar distributions of the dSphs are typically fit with either  
Plummer or a King profiles \cite{King:1962wi}. The primary difference
between  these fits is that Plummer profiles are described by a  
single parameter, $r_P$, and fall-off as a power law in the outer 
regions of the galaxy, while King profiles are described by a core
radius,  $r_K$, and a cut-off radius, $r_{cut}$, and fall-off exponentially   
in the outer regions. For the stellar distributions of Ursa Major II
and Coma,  we use the Plummer fits compiled in \cite{Simon:2007dq},
and for Willman 1 we use the King profile fit from
\cite{Martin:2007ic}. These quantities, along with the 
distance to each galaxy and their respective luminosities, are given in 
Table~\ref{tab:plummerking}. 

\begin{table*}
\begin{ruledtabular}
\begin{tabular}{lccccc}
  dSph   &  Distance (kpc) & Luminosity ($10^3$ L$_\odot$) & Core Radius (kpc)  
  & Cut-off Radius (kpc) & Number of stars \\
\hline
Ursa Major II   & 32 &2.8& 0.127 (P)&  ---&20 \\
Coma Berenices  & 44 &2.6& 0.064 (P) &  ---&59\\
Willman 1   & 38 &0.9& 0.02 (K) & 0.08 (K)& 47 \\
Ursa Minor & 66 & 290& 0.30 (K) & 1.50 (K) &187\\   
\end{tabular}
\end{ruledtabular}
\caption{ \label{tab:plummerking} The distance, luminosity, core 
and cut-off radii (Plummer (P), King (K)) for each of the dSphs we study. 
The last column gives the total number of stars used in the analysis.}
\end{table*}

We model the distribution of dark matter in the dSphs with radial
density profiles of the form
\be
\rho(r) = \frac{\rho_s}{ \tilde{r}^\gamma (1 + \tilde{r} ) ^{3-\gamma}}, 
\end{equation}
where $\rho_s$ is the characteristic density, $\tilde{r}=r / r_s$, and $r_s$ is the scale radius. 
Numerical simulations bound $\gamma$ in the range $[0.7-1.2]$,  and
the outer slope  is $\sim -3$ \cite{slopes}. 
In order to compare the mass distribution in dSphs to dark matter
halos in N-body simulations,  it is useful to work in terms of the two
parameters $\Vmax$ and $\rmax$, the maximum circular velocity  
and the radius at which it is obtained. For example, for  $\gamma =
(0.8,1,1.2)$,  $\rmax/r_s = (2.61,2.16,1.72)$ . 
Thus, for a particular value of $\gamma$, the density profiles of dark
matter halos can be described by either $\rho_s$-$r_s$ or similarly by
$V_{\rm max}$-$r_{\rm max}$.  

With the parameters of the halo density profile specified, the
$\gamma$-ray  flux from dark matter annihilations is given by 
\begin{equation}
\Phi = \frac{1}{2} P \int_0^{\xi_{max}} \sin \xi d \xi \int_{{\eta_{-}}}^{{\eta_{+}}} 
\left[ \frac{\rho_s}{\tilde{r}^\gamma ( 1 + \tilde{r})^{3-\gamma}} \right]^2 d \eta, 
\label{eq:gammarayflux}
\end{equation}
where $\eta_\pm = D \cos \xi \pm \sqrt{r_t^2 - D^2 \sin^2 \xi}$,  $D$
is the distance to the galaxy, $\xi$ is the angular distance from the
center of the galaxy,  and $r_t$ is the tidal radius for the dark
matter halo. 
Note that in the limit of $D \gg r_s$, the flux
scales as $\int \rho(r)^2 dr/D^2$, and in the particular case where $\gamma = 1$, 
$\Phi \sim \rho_s^2 r_s^3/D^2$, with $\sim 90\%$ of the flux originating within $r_s$.

In Eq.~(\ref{eq:gammarayflux}), the properties of the
dark matter particle are determined by 
 \be P= \frac{\langle  \sigma v \rangle}{M_x^2}  \int_{\Eth}^{\Mchi}
\frac{dN}{dE} dE.
\label{eq:Psusy}
\ee 
Here, $\Eth$ is a threshold energy, $\Mchi$ is the mass of the dark matter particle, 
$\langle \sigma v \rangle$ is the annihilation cross section, and the spectrum of 
the emitted $\gamma$-rays is given by $dN/dE$. Unless otherwise noted,  we assume 
$P \approx 10^{-28} {\rm cm^3} {\rm s}^{-2} {\rm GeV}^{-2}$, which corresponds to the most optimistic 
supersymmetric dark matter models. However, we stress that the derived results can be rescaled to any 
dark matter model, by a simple rescaling of $P$. 

\section{Likelihood Function and Priors}
\label{sec:likelihoodsection}
Observed stellar line-of-sight velocities place strong constraints on  several 
important parameters describing the dark matter halos of dSphs. Two  examples 
of these parameters are the halo mass and density at a characteristic halo radius, 
corresponding to about twice the King core radius (or about 600 pc for a typical 
dSph \cite{Walker:2007ju,Strigari:2007ma}). More relevant to $\gamma$-rays is the 
quantity $\rho_s^2 r_s^3$ (the $\gamma$-ray luminosity is ${\cal L} \sim \rho_s^2 r_s^3$), 
which is typically determined to within a factor $3-6$ with the line-of-sight velocities of 
several hundred stars \cite{Bergstrom:2005qk,Strigari:2006rd}. The constraints on these 
parameters can be strengthened by including relations between similar parameters 
observed in numerical simulations. When combined with the observational constraints, 
these empirical relations in numerical simulations constitute a theoretical prior, delineating 
a preferred region of the parameter space of the dark matter distribution in dSphs 
\cite{Strigari:2006rd,Penarrubia:2007zz}. In this section, we discuss the implementation of 
this prior, and derive the general form of the likelihood function we use to constrain the 
$\gamma$-flux from dark matter annihilations.

We assume the line-of-sight velocities are drawn from a Gaussian
distribution, centered on the true value of the mean velocity,
$u$. This has been shown to be a good description of  the well-studied
dwarfs with line-of-sight velocities of several hundred stars \cite{Walker:2005:nt}.
Given the set of theoretical parameters, the probability to obtain the set of observed 
line-of-sight velocities, $\vec{\bf x}$, is 
\be
P({\bf \vec{x}}|{\vec{\bf \theta}} ) 
= \prod_{i=1}^n 
\frac{1}{\sqrt{2 \pi ( \sigma_{t,\imath}^2 + \sigma_{m,\imath}^2})}
\exp \left [ - \frac{1}{2}\frac{(v_\imath - u)^2}{\sigma_{t,\imath}^2 
+ \sigma_{m,\imath}^2} \right ], 
\label{eq:likelihood} 
\ee 
Here $\vec{\theta}$ is the set of parameters that describe the dSph, and
the sum is over the observed total number of stars.
The dispersion in the velocities thus has two sources:  1) the intrinsic  
dispersion, $\sigma_{t,\imath}(\vec{\theta})$, which is a function of the position
of the $\imath^{th}$ star,  and 2) the uncertainty stemming from the
measurement, $\sigma_{m,\imath}$.

We can simplify Eq.~(\ref{eq:likelihood}) by assuming that the
measurement uncertainties are small  relative to the intrinsic 
dispersion.  This is a good approximation for well-studied dwarfs,
which have intrinsic dispersions  $\sim 10 \, {\rm km} \, {\rm
s}^{-1}$, and measurement errors $\sim 1 \, {\rm km} \, {\rm s}^{-1}$
\cite{dwarfdata}. Under this approximation Eq.~(\ref{eq:likelihood})
becomes
\begin{equation}
P({\bf \vec{x}}|{\vec{\bf \theta}}) =
\prod_{\imath=1}^{N_B}
\frac{1}{\sqrt{2\pi\sigma_{t,\imath}^2}}
\exp\left[-\frac{1}{2}
\frac{N_\imath\hat{\sigma}_{t,\imath}^2}{\sigma_{t,\imath}^2}\right]\,,
\label{eq:bin-approx} 
\end{equation}
where the sum is now over the number of bins, $N_B$, for which the
velocity dispersion  is determined. 
The velocity dispersion in the $\imath^{th}$ bin is $\hat{\sigma}_{t,\imath}^2$, and
the number of stars in the $\imath^{th}$ bin is 
$N_\imath$. We can use Eq.~({\ref{eq:bin-approx}) if the 
observations are given by line-of-sight velocity dispersions, 
and if the measurement errors are small in comparison to 
the intrinsic dispersion. This is the case for Ursa Minor, as discussed
below. 

We describe the dark matter halos in terms of the parameters 
$\vec{\theta} = (\Vmax, \rmax,\beta$). We assume that $\beta$ is constant
as a function of radius, and let it vary over the range $[-5:1]$. 
We integrate Eqs.~(\ref{eq:likelihood}) and (\ref{eq:bin-approx}) over 
these parameters and 
define the likelihood function for a fixed $\gamma$-ray flux, $f$,  as   
\bea  
{\cal L}(f)
&\propto& \int P({\bf \vec{x}} | V_{\rm max}, r_{\rm max},\beta)
P(V_{\rm max} , r_{\rm max})  \nonumber \\ &\times& \delta (\Phi(\Vmax,\rmax)-f)
dV_{\rm max} d r_{\rm max} d \beta.
\label{eq:Lflux}
\eea 
Here we have assumed a uniform prior on $\beta$, and  
the prior probability distribution for $V_{\rm max} , r_{\rm max}$ is 
given by $P(V_{\rm max} , r_{\rm max})$. This prior distribution is 
determined by the $\Vmax$-$\rmax$ relation from CDM simulations. 

In order to determine $P(V_{\rm max} , r_{\rm max})$,  we need both its mean
relation and its halo-to-halo scatter.
For dark subhalos that have been strongly affected by tidal interactions, 
the $V_{\rm max} - r_{\rm max}$ relation is strongly
dependent on the nature  of the potential of the host system, as
these systems undergo varying amounts of mass loss  as they evolve
within the host halo. For example, Bullock and Johnston~\cite{BJ07} 
have embedded a disk potential in
a MW size host halo, and found that the $\Vmax-\rmax$ relation of subhalos 
takes the form $\log (r_{\rm max}) = 1.35 [
\log (V_{\rm max}) -1] - 0.196$.  We obtain a similar slope by
examining the subhalos in the dark matter-only Via Lactea  simulation
of MW substructure \cite{Diemand:2006ik}, however differences
in the assumed  cosmological parameters, and the absence of a disk
potential in Via Lactea, translate into  differences in the
normalization of the $V_{\rm max}-r_{\rm max}$ relation. 
For Via Lactea subhalos, we find  the normalization is reduced by $\sim
30\%$, which implies reduced halo concentrations (larger $r_{\rm max}$
for fixed $V_{\rm max}$).

We model the scatter in  $V_{\rm max}$ as a log-normal distribution,
with $\sigma_{\log V_{\rm max}} \simeq 0.20$. This
provides a conservative  estimate for the scatter in $V_{\rm max}$ as
a function of $r_{\rm max}$ for nearly the entire  range of the
subhalo mass function. At the extremely high end of the subhalo mass function 
($V_{\rm max} \gtrsim 20 \, {\rm km} \, {\rm s}^{-1}$),  the scatter
increases because in this range it is dominated by a small
number of very massive systems that have been accreted into the host
halo very recently. 
This increase  in the scatter is simply because
Via Lactea is only one realization of a substructure population in a MW halo.
We find that by excluding the extreme
outliers in the Via Lactea mass function, the scatter is similar to
the low mass regime. This is a similar result to those obtained in semi-analytic 
models of many realizations of the subhalo population 
\cite{Zentner:2003yd,van den Bosch:2004zs}.

\begin{table}
\begin{ruledtabular}
\begin{tabular}{lcc}
  dSph   &  Mass $<$ 100 pc [$10^6$ M$_\odot$] & V$_{\rm max}$ [km s$^{-1}$]\\
\hline
Ursa Major II    & $3.1_{-1.8}^{+5.6}$& $23_{-10}^{+69}$\\
Coma Berenices & $1.9_{-1.0}^{+2.1}$&$19_{-9}^{+53}$ \\
Willman 1   & $1.3_{-0.8}^{+1.5}$ &$27_{-15}$ \\
Ursa Minor & $2.3_{-1.2}^{+1.9}$& $30_{-16}^{+12}$\\
\end{tabular}
\end{ruledtabular}
\caption{\label{tab:masstable}
The masses within 100 pc and the maximum circular velocities
of the Milky Way satellites we study. Error bars indicate the $90\%$ c.l. regions. 
No upper limit could be obtained for the maximum circular velocity of 
Willman 1. 
 }
\end{table}


\begin{figure}[hbtp]
\begin{center}
\includegraphics[height=8.cm]{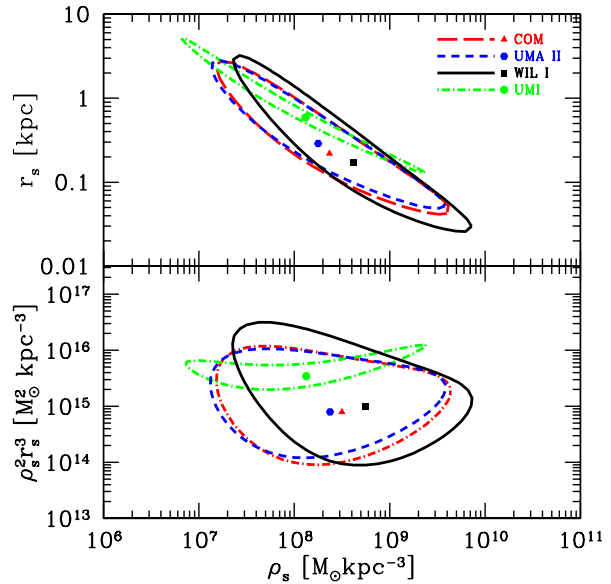}  
\caption{\small The $90\%$ confidence level 
region in the $\rho_s-r_s$ (top) $\rho_s^2 r_s^3-\rhos$ (bottom) parameter space 
for Coma , Ursa Major II, Willman 1, and Ursa Minor. 
We marginalize over the velocity anisotropy and have assumed an 
inner slope of $\gamma =1$. The best-fit values are indicated with points. 
\label{fig:rhosrs}}
\end{center}
\end{figure}

\section{Results}
\label{sec:results}
\subsection{Flux estimates for smooth dark matter distributions}

We now quantify the prospects for detecting $\gamma$-rays from dark matter 
annihilation in the three new dSphs. We first assume that the dark matter is distributed 
smoothly, and we discuss the implications of a boost factor due to substructure in the 
next subsection. We use observations of these galaxies from the following references: 
Ursa Major II and Coma  \cite{Simon:2007dq}, Willman 1 \cite{Willmaninprep}. For these 
galaxies we have individual stellar velocities, so we use the likelihood function in 
Eq.~(\ref{eq:likelihood}). To make a connection to previous studies of dSphs, we compare 
the fluxes for these new dSphs to the flux from Ursa Minor, which is at a distance $D=66$ 
kpc, and has a luminosity of $L=2.9 \times 10^5$ L$_\odot$. Ursa Minor has the largest flux of 
any of the well-known dwarfs \cite{Strigari:2006rd}. We describe the stellar distribution of 
Ursa Minor with a King profile, with $r_K = 0.30$ kpc and $r_{cut} =1.50$ kpc 
\cite{Munoz:2005be}.  For Ursa Minor we use the measured velocity dispersion from a 
sample of 187 stars distributed evenly in 11 bins \cite{Munoz:2005be}, and we use the 
likelihood function in Eq.~(\ref{eq:bin-approx}). 

\begin{figure}[hbtp]
\begin{center}
\includegraphics[height=8cm]{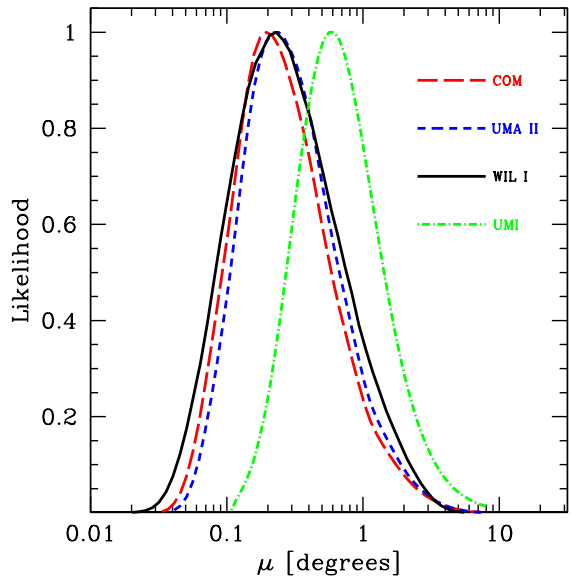}   
\caption{\small The probability distributions of the angular size subtended by $r_s$ 
for each galaxy. We 
marginalize over the velocity anisotropy and $\rho_s$. The inner 
slope is fixed to $\gamma=1$. 
\label{fig:rs}}
\end{center}
\end{figure}

In the top panel of Fig.~(\ref{fig:rhosrs}), we show the $90\%$ c.l. 
region in the $\rhos -\rs$ plane for each galaxy, where the best-fit 
values are denoted by points. Here we use the $\Vmax-\rmax$ prior from 
section~\ref{sec:likelihoodsection}, and we take the inner slope of the dark matter
halo profile to be $\gamma = 1$. In the bottom panel 
of Fig.~(\ref{fig:rhosrs}) we show the $90\%$ confidence level region in the 
$\rhos^2 \rs^3-\rhos$ plane. As seen in Fig.~(\ref{fig:rhosrs}), the range of 
values that $\rhos^2 \rs^3$ can take in each dSph is reduced with the inclusion 
of more stars (e.g. Ursa Minor vs any one of the other three dSphs). 

The constrained regions in Fig.~(\ref{fig:rhosrs}) can be used to determine the
masses, maximum circular velocities, and $\gamma$-ray flux 
probability distributions for each galaxy. In
Table~\ref{tab:masstable}, we show  
the masses within 100 pc and the maximum circular velocities for each galaxy. The 
error bars indicate the $90\%$ c.l. regions. 

To determine the flux distributions, we must 
first specify a solid angle for integration. For optimal detection
scenarios, the solid angle should  encompass the region with the
largest signal-to-noise. For the present work, we will integrate over
a region where 90\% of the flux originates. As discussed above, for
the particular case where 
$\gamma=1$, 90\% of the flux originates  within $\rs$. Therefore, in
order to estimate the solid angle of  integration, we have to
first determine the maximum likelihood values of $\rs$. This is  done
by marginalizing over $\rhos$ and $\beta$ with the $\Vmax - \rmax$
prior.  The distributions of angular sizes are then obtained from   
$\mu = \tan^{-1}[ \rs / D]$, where $D$ is the distance to the dSph.  
As shown in Fig.~(\ref{fig:rs}), we find that given their similar 
size and roughly similar distances, all three dSphs will emit 90\% of 
their $\gamma-$ray flux within a region of $\sim 0.2$ degrees,
centered on each dSph (for $\gamma=1$).  
Ursa Minor is the most physically extended galaxy, subtending the
largest projected area on the sky.  

\begin{figure}[hbtp]
\begin{center}
\includegraphics[height=8cm]{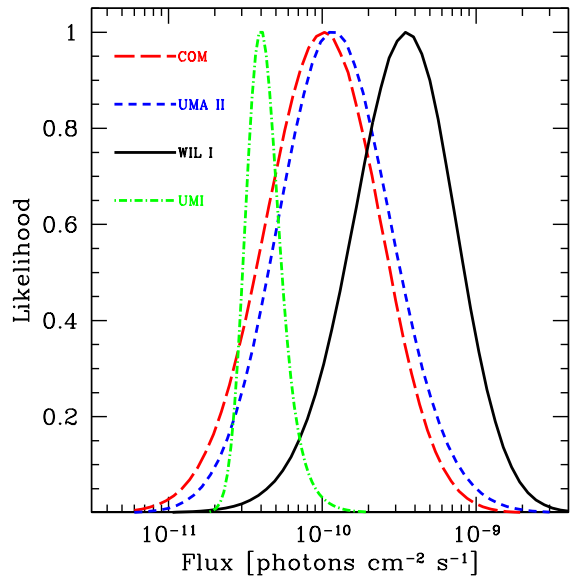}  
\caption{\small The probability distributions for the $\gamma$-ray fluxes 
from Coma, Ursa Major II, Willman 1, and Ursa Minor, marginalizing over 
the velocity anisotropy, $\rho_s$, and $r_s$. We assume 
$P= 10^{-28} \, {\rm cm}^{3} \, {\rm s}^{-1} \, {\rm GeV}^{-2}$ and an inner 
slope of $\gamma = 1.0$. We have assumed no boost from 
halo substructure, which increases these fluxes by a factor $\sim 10-100$.  
}
\label{fig:flux}
\end{center}
\end{figure}

It is important to determine whether each of the galaxies will be detected as point sources, 
or whether they will be resolved as extended objects. To determine this we compare their 
angular size to the angular resolution of $\gamma$-ray telescopes. GLAST will have a single 
photon angular resolution of $\sim 10$ arcminutes for energies greater than 1 GeV, similar to 
the angular resolution of ground-based detectors (such as VERITAS) for energies greater than 
few tens of GeV. In the case where the detected number of photons is $N_\gamma > 1$, the 
angular resolution of a detector is improved by a factor of $1 / \sqrt{N_\gamma}$. Therefore, 
these galaxies can be resolved as extended objects, which in principle would allow a measured 
flux to determine the distribution of dark matter in the halo itself. 

Fig.~(\ref{fig:flux}) depicts the resulting flux probability distribution for the three new 
dSphs and Ursa Minor. These 
are obtained by marginalizing over $\beta$, $\rho_s$, and $r_s$ and including 
the $V_{\rm max}$-$r_{\rm max}$ prior. 
We set the inner slope to $\gamma = 1$, and integrate the flux over the solid angle 
that corresponds to 0.2 degrees from the center of the galaxy. We assume a value of 
$P=P_0=10^{-28} \cm^3 \s^{-1} \GeV^{-2}$, but the result can be scaled to any 
dark matter candidate with a different value of $P$ 
by simply multiplying the flux distribution by a factor of $P/P_0$. 

\begin{figure}[hbtp]
\begin{center}
\includegraphics[height=5.cm]{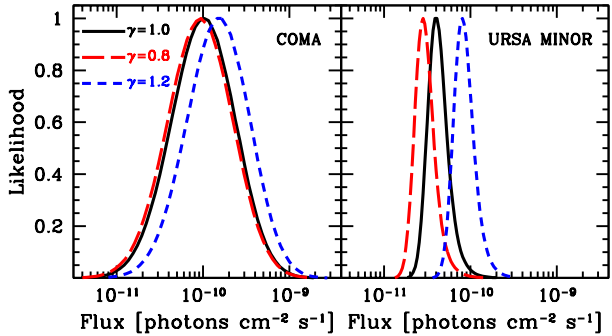}
\caption{\small The $\gamma$-ray flux probability distributions for Coma  and Ursa Minor 
for inner slopes of 0.8 (long-dashed), 1.0 (solid), and 1.2 (short-dashed). We marginalize over 
the same quantities as in Fig.~(\ref{fig:flux}). The value of $P$ is the same as in Fig.~(\ref{fig:flux}).
\label{fig:slope}
}
\end{center}
\end{figure}

The relative proximity of the three new dSphs, and their comparable sizes, results in 
$\gamma$-ray fluxes that are roughly similar. For $P\approx P_0$, the likelihood peaks at 
approximately $\Phi_0 \approx 10^{-10} \cm^{-2} \s^{-1}$, with a spread of nearly an order of 
magnitude. Thus Ursa Major II, Coma, and Ursa Minor all have comparable fluxes, 
and Willman 1 has a most likely flux that is about three times larger than Ursa Major II 
or Coma. 


\subsection{The effects of the inner slope and substructure boost factors}

Understanding the distribution of dark matter in the inner regions of the dSphs also 
has important implications for detection of a $\gamma$-ray flux. 
However, when varying the inner slope, we must also be certain to vary all of the other
halo parameters so as to remain consistent with the line-of-sight observations. 
In order to quantify the effects of varying the inner slope, we marginalize over $\Vmax$, $\rmax$ and $\beta$ for 
profiles with different values of $\gamma$. In Fig.~(\ref{fig:slope}) we show the effects of 
varying $\gamma$ for Coma  and Ursa Minor. The shifts in the flux distribution function are 
not only a result of varying the inner slope but also come from the constraints imposed 
by the data on the density profile parameters $\rhos$ and $r_s$.  The relative amount of
the shifts can be understood by considering the best-fitting values of $r_s$ (where the
majority of the $\gamma$-ray flux comes from) as compared to the core radii of the 
systems. When the core radius is similar to $r_s$, the shifts are larger for varying 
$\gamma$, as in the case of Ursa Minor. However, when the core radius is much smaller 
than the fitted values of $r_s$, variations in the inner slope are less significant than the 
change induced by $\rhos$, as is the case for Coma. 

The presence of substructure in dark matter halos is firmly established on theoretical and 
numerical grounds. Dark matter halos are approximately self-similar, and substructure is 
expected to be present in all dark matter halos with mass greater than the cut-off scale 
in the primordial power spectrum, set by the kinetic decoupling temperature of the 
dark matter particle (for detecting the smallest dark matter halos in the Milky Way, see 
\cite{Koushiappas:2006qq}). It is therefore natural to expect that these galaxies contain 
substructure if they consist of CDM. 

The density enhancement over the smooth distribution of dark matter leads to an 
enhancement in the total annihilation rate, typically quantified in terms of a ``boost" factor. 
As was shown in \cite{Strigari:2006rd}, the boost factor cannot attain arbitrarily large values, 
but instead is bounded to be less than $\sim 100$, with the exact value depending on the 
cut-off scale in the CDM mass function. 
The boost is a multiplicative quantity, so the effect of dark substructure is simply 
accounted for by scaling the fluxes in Fig.~(\ref{fig:flux}) by the appropriate boost factor. 

\subsection{Detection prospects}

As is shown in Fig.~(\ref{fig:flux}), the flux probability
distribution functions peak around  
$\Phi_0 \approx 10^{-10} \cm^{-2} \s^{-1}$ without including any
enhancement to the  signal from substructure. Here, we will assume a
conservative value of 10 for the  boost factor, and discuss the
prospects for detecting the three new dSphs with $\gamma$-ray
instruments.  We can make simple estimates for the likelihood for
detection by adopting the specifications  
of particular $\gamma$-ray detectors. We will use two examples: a
space-based experiment,  
GLAST, and a ground-based Cerenkov telescope, VERITAS. 
For GLAST, if we assume an orbit-averaged effective area of $\Aeff
\approx 2 \times 10^3 \cm^2$  
and an exposure time of $\texp = 10$ years, their product is 
$B_{\rm G } = \Aeff \texp \approx 3 \times 10^{11} \cm^2 \s$.  A 50
hour exposure with  VERITAS ($\Aeff \approx 10^8 \cm^2$) has $B_{\rm
  V} \approx  2 \times 10^{13} \cm^2 \s$.  
Naively, for a fixed value of $P$, a ground-based detector seems more
sensitive because $\BV > \BG$. However, the backgrounds for a
ground-based detector are also larger and include a component from
the hadronization of cosmic rays in the atmosphere.

As an example, for a fixed value of $ P = P_0$, and a solid angle
that  corresponds to an angular size of 0.2 degrees, the number of
photons detected  by GLAST is $N_{\gamma, G} \approx 300$. The
dominant source of background  for GLAST is the Galactic diffuse
emission ($dN_B /dE = 1.2 \times 10^{-6} [\Eth/ \GeV]^{-1.1} \cm^{-2}  
\s^{-1} {\rm sr}^{-1} \GeV^{-1}$ \cite{Hunter:1997we}). If we assume an
energy threshold  
of $\Eth \approx 1 \GeV$, then the number of background photons is 
$N_B = B_{\rm G} \, dN_B/ dE \approx 250$, which means that the new satellites will 
be detected at approximately a $N_{\gamma,G} / \sqrt{N_B+N_{\gamma,G}}  
\approx 12 \sigma$ level. A similar estimate can be obtained for
VERITAS. The number  
of photons detected above $50\GeV$ in an instrument with an effective
area times exposure $\sim B_{\rm V}$ is  
$N_\gamma \approx 2 \times 10^4$. The dominant contribution to the
background are  photons that originate from neutral pion decays from
the nuclear interactions of cosmic  rays in the upper layers of the
atmosphere  
($dN_B/dE = 3.8 \times 10^{-3} [\Eth / \GeV]^{-2.75} \cm^{-2} \s^{-1}
{\rm sr}^{-1} \GeV^{-1}$  
\cite{Nishimura:1980pz}). The number of background photons from pion decays is 
approximately $N_B \approx 2 \times 10^7$, therefore the 3 dSphs could be detected at 
a $\nu \approx 5 \sigma$ level. Understanding and discriminating against the background 
contamination in Cerenkov telescopes is very important in improving the prospects for 
detecting dSphs of the Milky Way. 

As shown in \cite{Strigari:2006rd}, the large number of stellar
velocities obtained in the older dSphs allow useful $\gamma$-ray
flux ratios between different dSphs to be determined. For the 
three new dSphs considered in this work, the kinematic  
data is not good enough to play the same game. Clearly, more stellar
velocities will shrink the allowed region of $\rhos^2 \rs^3$ parameter
permitting robust estimates of flux ratios between the galaxies
studied here and the rest  of the Milky Way satellites.  

\section{Discussion \& Conclusions}
\label{sec:conclusions}
In this paper, we have modeled the dark matter distribution in three recently-discovered 
Milky Way (MW) satellites (Ursa Major II, Willman 1, and Coma Berenices), and have presented 
the prospects for detecting $\gamma$-rays from dark  matter annihilations in their halos.  
We show that the expected flux from these galaxies 
is larger than the flux from any of the higher luminosity, more well-known (pre-SDSS)
dwarfs. There are two reasons for this surprising result: 1) the masses of
these new dwarfs within their stellar distributions are similar to the
masses of the well-known, larger luminosity  dwarfs, and 2) all 
three new galaxies are closer than the other well-known
dwarfs. The implied mass-to-light ratios, $\sim 1000$, of these new
dwarfs  makes them the most dark matter dominated galaxies in the
Universe.  

Our estimates show that it is unlikely that the observed stellar
distributions are presently undergoing tidal disruption. However,
this does not mean that they have been free from tidal interactions in
the past, but  rather that the surviving stellar core can be
faithfully modeled as a system in 
dynamical  equilibrium. By including the $\Vmax-\rmax$ CDM prior, we
have naturally accounted for tidal effects in the mass modeling, since
this $\Vmax-\rmax$ relation in fact comes  from dark matter halos that
have experienced tidal stripping.  

One of the galaxies we consider, Ursa Major II, may be a candidate for
past tidal disruption,  given that it is positioned on the same great
circle as the Orphan Stream of stars, which was also recently detected by
the SDSS~\cite{Zucker:2006bf,Belokurov:2006ph}.  
This is consistent with the findings of Simon and Geha
\cite{Simon:2007dq}, who have recently investigated the possibility of
tidal disruption in Ursa Major II, as well as all of the  other new
dwarfs, using proxies such as gradients in the observed velocity
distribution and metallicity of the stellar populations. Given the
total mass-to-light  ratio we have determined for Ursa Major II, tidal
stripping will have only been significant if its pericenter is $\sim$
3 times closer than its  present distance. Future observations of the
stellar distributions in Ursa Major II, and all of the other new   
faint dwarfs, will be important in determining bound and unbound
stellar populations.  With a larger sample of stars from a galaxy such
as Ursa Major II, it will be possible to  remove unbound and
interloping stars with  techniques similar to those presented in
\cite{tides}. Upon removal of  stars unbound to the galaxy, these
authors show that in most cases the true bound  mass of the system can
be recovered to typically better than $25\%$. 

As a very conservative check for the effects of membership
uncertainties, we have redone the analysis for each of the galaxies by
just keeping the stars within the inner half of each galaxy, where the
surface densities are the most well-determined. We find that the peak
of the  flux likelihood is shifted by a small amount relative to the
1-$\sigma$ widths in Fig.~(\ref{fig:flux}). Note however, that even
when including the entire population of stars in the observed
samples, in all cases equilibrium models provide adequate descriptions
of the dynamics of each system. 

Unresolved binary star systems also introduce a systematic that may
effect the fluxes we have presented.  Olszewski, Pryor, and Armandroff
\cite{Olszewski:1995gs} have determined the effect of binaries on the
velocity dispersion of two of the most luminous dwarfs, Draco and Ursa
Minor, by inferring the binary population of these systems  using
multiple epoch observations. They find a velocity dispersion of  $\sim
1.5 \, {\rm km} \, {\rm s}^{-1}$ due to binaries, and a probability of   
$5 \%$ that binaries elevate the velocity dispersion to $4 \, {\rm km}
\, {\rm s}^{-1}$,  
which is still less than the velocity dispersion of the three new dwarfs. 
Thus if Ursa Major II, Willman 1, and Coma have binary fractions
similar to Draco and Ursa Minor, their observed velocity dispersions
are not significantly affected by  binaries. We note that this is
consistent with recent estimates of the binary fraction in low density
Galactic globular clusters \cite{Sollima:2007sc}.  

A strong test for the presence of binaries is to examine the
distribution of measured velocities. The velocity
distribution due to internal motion in binary systems should be flat
due to the observed broadness of the period distribution of binaries  
\cite{Duquennoy:1991zu}. The observed period distributions depend
on spectral type and age of the system (among other variables) but are
all broad with dispersions of about two orders of magnitude, which
seems to be consistent with theoretical expectations
\cite{Fisher:2003tg}. Further, there is no observational evidence or 
theoretical argument that suggests that the period distribution
should be sharply suppressed for all periods below $\sim 1000$ years 
(roughly velocities larger than $\sim 3$ km/s). Thus if there is a
large contribution from internal motion in binary stellar systems to
the intrinsic velocity dispersion of these dwarfs, we expect to see a
significant tail of high velocities. This is not observed and hence we
can be confident that measured velocity  dispersion is tracing the
total mass in the dwarf galaxy.  

The new, ultra-low luminosity galaxies represent an interesting
confluence of astronomical and $\gamma$-ray studies.  
Future kinematic studies of all of these new dwarfs will 
further reduce  the errors on the mass distributions, and sharpen the
predictions for $\gamma$-ray observatories searching for signatures of
dark matter annihilations. 


\section{Acknowledgments} 
We are grateful to Simon White for many useful discussions on likelihood 
functions for dwarf galaxies; Jay Strader and Connie Rockosi for
Willman 1 data;    Juerg Diemand, Michael Kuhlen, and Piero Madau for
making the data from  the  Via Lactea simulation publicly available;
and John Beacom, Nicolas Martin, and  Joe Wolf for useful
discussions. JSB, LES, and MK are supported in part by  NSF  grant
AST-0607746. Work at LANL was carried out under the auspices of the
NNSA  of the U.S. Department of Energy at Los Alamos National
Laboratory under  Contract No.  DE-AV52-06NA25396. JDS gratefully
acknowledges the support of a  Millikan Fellowship provided by
Caltech.


\begin{thebibliography}{99}

\bibitem{Mateo:1998wg}
  M.~Mateo,
  Ann.\ Rev.\ Astron.\ Astrophys.\  {\bf 36}, 435 (1998)
  [arXiv:astro-ph/9810070].

\bibitem{discovery}
  B.~Willman {\it et al.},
  Astrophys.\ J.\  {\bf 626}, L85 (2005)
  [arXiv:astro-ph/0503552]; 
  M.~J.~Irwin {\it et al.},
  Astrophys.\ J.\  {\bf 656}, L13 (2007)
  [arXiv:astro-ph/0701154];
  S.~M.~Walsh, H.~Jerjen and B.~Willman,
  arXiv:0705.1378 [astro-ph]. 

\bibitem{Zucker:2006bf}
  D.~B.~Zucker {\it et al.},
  Astrophys.\ J.\  {\bf 650}, L41 (2006)
  [arXiv:astro-ph/0606633]. 
    
\bibitem{Belokurov:2006ph}
  V.~Belokurov {\it et al.}  [SDSS Collaboration],
  Astrophys.\ J.\  {\bf 654}, 897 (2007)
  [arXiv:astro-ph/0608448]. 
  
\bibitem{missingsatellites}
  G.~Kauffmann, S.~D.~M.~White and B.~Guiderdoni,
  Mon.\ Not.\ Roy.\ Astron.\ Soc.\  {\bf 264} (1993) 201.
  A.~A.~Klypin, A.~V.~Kravtsov, O.~Valenzuela and F.~Prada,
  Astrophys.\ J.\  {\bf 522}, 82 (1999)
  [arXiv:astro-ph/9901240].
  B.~Moore, S.~Ghigna, F.~Governato, G.~Lake, T.~Quinn, J.~Stadel and P.~Tozzi,
  Astrophys.\ J.\  {\bf 524}, L19 (1999).
  
\bibitem{dsp}  
  J.~S.~Bullock, A.~V.~Kravtsov and D.~H.~Weinberg,
  Astrophys.\ J.\  {\bf 539}, 517 (2000)
  [arXiv:astro-ph/0002214];
W.   A.   Chiu,   N.   Y.   Gnedin,  and   J.   P.   Ostriker,
Astrophys.   J.   {\bf   563},    21   (2001) [astro-ph/0103359];   
  A.~J.~Benson, C.~G.~Lacey, C.~M.~Baugh, S.~Cole and C.~S.~Frenk,
  Mon.\ Not.\ Roy.\ Astron.\ Soc.\  {\bf 333}, 156 (2002)
  [arXiv:astro-ph/0108217];
R.  Barkana and A. Loeb, Astrophys.  J. {\bf 523},
54 (1999) [arXiv:astro-ph/9901114]; 
A.  Dekel and J. Silk, Astrophys.  J. {\bf 303},
38  (1986);  
  S.~Cole, A.~Aragon-Salamanca, C.~S.~Frenk, J.~F.~Navarro and S.~E.~Zepf,
  Mon.\ Not.\ Roy.\ Astron.\ Soc.\  {\bf 271}, 781 (1994)
  [arXiv:astro-ph/9402001];
  R.~S.~Somerville and J.~R.~Primack,
  Mon.\ Not.\ Roy.\ Astron.\ Soc.\  {\bf 310}, 1087 (1999)
  [arXiv:astro-ph/9802268].

\bibitem{Walker:2007ju}
  M.~G.~Walker, M.~Mateo, E.~W.~Olszewski, O.~Y.~Gnedin, X.~Wang, B.~Sen and M.~Woodroofe,
  arXiv:0708.0010 [astro-ph].
  
\bibitem{Strigari:2007ma}
  L.~E.~Strigari, J.~S.~Bullock, M.~Kaplinghat, J.~Diemand, M.~Kuhlen and P.~Madau,
  arXiv:0704.1817 [astro-ph].

\bibitem{Martin:2007ic}
  N.~F.~Martin, R.~A.~Ibata, S.~C.~Chapman, M.~Irwin and G.~F.~Lewis,
  arXiv:0705.4622 [astro-ph].
  
\bibitem{Simon:2007dq}
  J.~D.~Simon and M.~Geha, Astrophys.\ J.\ in press, 
  arXiv:0706.0516 [astro-ph].
  
\bibitem{GLAST} S.~ Ritz., J.~Grindlay, 
C.~Meegan, P.~F.~Michelson,  \& GLAST Mission Team 2005, Bulletin of the 
American Astronomical Society, 37, 1198. 

\bibitem{Hanna:2002bf}
  D.~S.~Hanna {\it et al.},
  Nucl.\ Instrum.\ Meth.\  A {\bf 491}, 126 (2002)
  [arXiv:astro-ph/0205510].
   
\bibitem{Hofmann:2003kx}
  W.~Hofmann  [HESS Collaboration],
{\it Prepared for 28th International Cosmic Ray Conferences (ICRC 2003), Tsukuba, Japan, 31 Jul - 7 Aug 2003}.

\bibitem{Cortina:2005pt}
  J.~Cortina {\it et al.}  [MAGIC Collaboration],
{\it Prepared for 29th International Cosmic Ray Conference (ICRC 2005), Pune, India, 3-11 Aug 2005}

\bibitem{Weekes:2001pd}
  T.~C.~Weekes {\it et al.},
  Astropart.\ Phys.\  {\bf 17}, 221 (2002)
  [arXiv:astro-ph/0108478].
  
\bibitem{Yoshikoshi:1999rg}
  T.~Yoshikoshi {\it et al.},
  Astropart.\ Phys.\  {\bf 11}, 267 (1999).

\bibitem{Sinnis:2005un}
  G.~Sinnis  [HAWC Collaboration],
  AIP Conf.\ Proc.\  {\bf 745} (2005) 234.
  
\bibitem{previousdwarfs}  
  E.~A.~Baltz, C.~Briot, P.~Salati, R.~Taillet and J.~Silk,
  Phys.\ Rev.\  D {\bf 61}, 023514 (2000)
  [arXiv:astro-ph/9909112]; 
  N.~W.~Evans, F.~Ferrer and S.~Sarkar,
  Phys.\ Rev.\  D {\bf 69}, 123501 (2004)
  [arXiv:astro-ph/0311145];
  S.~Profumo and M.~Kamionkowski,
  JCAP {\bf 0603}, 003 (2006)
  [arXiv:astro-ph/0601249];
  M.~A.~Sanchez-Conde, F.~Prada, E.~L.~Lokas, M.~E.~Gomez, R.~Wojtak and M.~Moles,
  arXiv:astro-ph/0701426.

\bibitem{Bergstrom:2005qk}
  L.~Bergstrom and D.~Hooper,
  Phys.\ Rev.\  D {\bf 73}, 063510 (2006)
  [arXiv:hep-ph/0512317]. 
  
\bibitem{Strigari:2006rd}
  L.~E.~Strigari, S.~M.~Koushiappas, J.~S.~Bullock and M.~Kaplinghat,
  Phys.\ Rev.\  D {\bf 75}, 083526 (2007)
  [arXiv:astro-ph/0611925].

\bibitem{galacticcenter}
  D.~Hooper and B.~L.~Dingus,
  Phys.\ Rev.\  D {\bf 70}, 113007 (2004)
  [arXiv:astro-ph/0210617].

\bibitem{microhalos} 
  C.~Calcaneo-Roldan and B.~Moore,
  Phys.\ Rev.\  D {\bf 62}, 123005 (2000)
  [arXiv:astro-ph/0010056];
  A.~Tasitsiomi and A.~V.~Olinto,
  Phys.\ Rev.\  D {\bf 66}, 083006 (2002)
  [arXiv:astro-ph/0206040];
  F.~Stoehr, S.~D.~M.~White, V.~Springel, G.~Tormen and N.~Yoshida,
  Mon.\ Not.\ Roy.\ Astron.\ Soc.\  {\bf 345}, 1313 (2003)
  [arXiv:astro-ph/0307026];
  S.~M.~Koushiappas, A.~R.~Zentner and T.~P.~Walker,
  Phys.\ Rev.\  D {\bf 69}, 043501 (2004)
  [arXiv:astro-ph/0309464];
  E.~A.~Baltz, J.~E.~Taylor and L.~L.~Wai,
  arXiv:astro-ph/0610731;
  L.~Pieri, E.~Branchini and S.~Hofmann,
  Phys.\ Rev.\ Lett.\  {\bf 95}, 211301 (2005)
  [arXiv:astro-ph/0505356];
\bibitem{Koushiappas:2006qq}
  S.~M.~Koushiappas,
  Phys.\ Rev.\ Lett.\  {\bf 97}, 191301 (2006)
  [arXiv:astro-ph/0606208].

\bibitem{Diemand:2006ik}
  J.~Diemand, M.~Kuhlen and P.~Madau,
  Astrophys.\ J.\  {\bf 657}, 262 (2007)
  [arXiv:astro-ph/0611370].
  
\bibitem{tides} 
S.~Piatek and C.~Pryor, \aj {\bf 109}, 1071 (1995);   
  J.~Klimentowski, E.~L.~Lokas, S.~Kazantzidis, F.~Prada, L.~Mayer and G.~A.~Mamon,
  Mon.\ Not.\ Roy.\ Astron.\ Soc.\  {\bf 378}, 353 (2007)
  [arXiv:astro-ph/0611296].
  
\bibitem{Penarrubia:2007zx}
  J.~Penarrubia, J.~F.~Navarro and A.~W.~McConnachie,
  arXiv:0708.3087 [astro-ph].

\bibitem{Lokas:2004sw}
  E.~L.~Lokas, G.~A.~Mamon and F.~Prada,
  Mon.\ Not.\ Roy.\ Astron.\ Soc.\  {\bf 363}, 918 (2005)
  [arXiv:astro-ph/0411694].

\bibitem{Mashchenko:2005bj}
  S.~Mashchenko, A.~Sills and H.~M.~P.~Couchman,
  Astrophys.\ J.\  {\bf 640}, 252 (2006)
  [arXiv:astro-ph/0511567].

\bibitem{Gilmore:2007fy}
  G.~Gilmore, M.~I.~Wilkinson, R.~F.~G.~Wyse, J.~T.~Kleyna, A.~Koch, N.~W.~Evans and E.~K.~Grebel,
  arXiv:astro-ph/0703308.
  
\bibitem{King:1962wi}
  I.~King,
  Astron.\ J.\  {\bf 67}, 471 (1962).
  
\bibitem{slopes}
  J.~F.~Navarro {\it et al.},
  Mon.\ Not.\ Roy.\ Astron.\ Soc.\  {\bf 349}, 1039 (2004)
  [arXiv:astro-ph/0311231];
  J.~Diemand, M.~Zemp, B.~Moore, J.~Stadel and M.~Carollo,
  Mon.\ Not.\ Roy.\ Astron.\ Soc.\  {\bf 364}, 665 (2005)
  [arXiv:astro-ph/0504215].
      
\bibitem{Penarrubia:2007zz}
  J.~Penarrubia, A.~McConnachie and J.~F.~Navarro,
  arXiv:astro-ph/0701780.
  
\bibitem{Walker:2005:nt} 
M.~G.~Walker, M.~Mateo, E.~W.~Olszewski, , R.~Bernstein, X.~Wang, 
\& M.~Woodroofe, \aj, {\bf 131} 2114 (2006)     
[arXiv:astro-ph/0511465]. 

\bibitem{dwarfdata}
K.~B.~Westfall  {\it et al}, \aj, {\bf 131} 375 (2006)
[arXiv:astro-ph/0508091];
  A.~Koch {\it et al.},
  Astrophys.\ J.\  {\bf 657}, 241 (2007)
  [arXiv:astro-ph/0611372].
    
\bibitem{BJ07} 
J.~S.~Bullock and K.~V.~Johnston, Island Universes, 
Astrophysics and Space Science  Proceedings~
Springer, ~227 (2007).   
    
\bibitem{Zentner:2003yd}
  A.~R.~Zentner and J.~S.~Bullock,
  Astrophys.\ J.\  {\bf 598}, 49 (2003)
  [arXiv:astro-ph/0304292].
  
\bibitem{van den Bosch:2004zs}
  F.~C.~van den Bosch, G.~Tormen and C.~Giocoli,
  Mon.\ Not.\ Roy.\ Astron.\ Soc.\  {\bf 359}, 1029 (2005)
  [arXiv:astro-ph/0409201].
  
\bibitem{Willmaninprep}
B.~Willman, M.~Geha, J.~Strader, and C.~Rockosi in preparation.
  
\bibitem{Munoz:2005be}
  R.~R.~Munoz {\it et al.},
  Astrophys.\ J.\  {\bf 631}, L137 (2005)
  [arXiv:astro-ph/0504035]. 
  
\bibitem{Hunter:1997we}
  S.~D.~Hunter {\it et al.},
  Astrophys.\ J.\  {\bf 481}, 205 (1997).
    
\bibitem{Nishimura:1980pz}
  J.~Nishimura {\it et al.},
  Astrophys.\ J.\  {\bf 238}, 394 (1980).

\bibitem{Olszewski:1995gs}
  E.~Olszewski, C.~Pryor and T.~Armandroff,
  Astron. J. {\bf 111}, 750 (1996)
  [arXiv:astro-ph/9510155].

\bibitem{Sollima:2007sc}
  A.~Sollima, G.~Beccari, F.~R.~Ferraro, F.~Fusi Pecci and A.~Sarajedini,
  arXiv:0706.2288 [astro-ph].
  
\bibitem{Duquennoy:1991zu}
  A.~Duquennoy and M.~Mayor,
  Astron.\ Astrophys.\  {\bf 248}, 485 (1991).

\bibitem{Fisher:2003tg}
  R.~T.~Fisher,
  Astrophys.\ J.\  {\bf 600}, 769 (2004)
  [arXiv:astro-ph/0303280].
  
       
\end{thebibliography}
\end{document}